# Ordered phases in coupled nonequilibrium systems: dynamic properties

Shauri Chakraborty(1), Sakuntala Chatterjee(1) and Mustansir Barma(2)
*(1) Department of Theoretical Sciences, S.N. Bose National Centre for Basic Sciences,*
*JD Block, Sector 3, Salt Lake, Kolkata - 700106, India*
*(2) TIFR Centre for Interdisciplinary Sciences, Tata Institute of*
*Fundamental Research, Gopanpally, Hyderabad 500107, India.*

We study the dynamical properties of the ordered phases obtained in a coupled nonequilibrium system describing advection of two species of particles by a stochastically evolving landscape. The local dynamics of the landscape also gets affected by the particles. In a companion paper we have presented static properties of different phases that arise as the two-way coupling parameters are varied. In this paper we discuss the dynamics. We show that in the ordered phases macroscopic particle clusters move over an ergodic time-scale growing exponentially with system size but the ordered landscape shows dynamics over a faster time-scale growing as a power of system size. We present a scaling ansatz that describes several dynamical correlation functions of the landscape measured in steady state.

# I. INTRODUCTION

The theory of non-linear fluctuating hydrodynamics has recently been used extensively to study the dynamics of coupled driven systems with more than one conserved field [1–3]. A basic underlying assumption of such analyses is that in steady state, the system is spatially homogeneous on a macroscopic scale; this allows one to write a hydrodynamic expansion of the local fields around their homogeneous, stationary values. The dynamics of the coupled modes which ensue can then be studied, and interesting results have been obtained for their speeds and decay times.

However, if the coupling between the conserved fields is such that it drives the coupled system to an ordered, phase separated state, the hydrodynamic theory described above fails: the amplitudes and extent of spontaneously formed density inhomogeneities grow in time, ultimately resulting in a phase separated state. While linear stability analysis based on hydrodynamics may help in identifying the threshold for instability, a different approach is required to understand the dynamic properties of the new steady state, as the questions of interest are now different.

In this paper, we study the steady state dynamics of a coupled nonequilibrium system which exhibits different types of phase separated states, as the coupling parameters are varied. The system is best thought of as two species of particles with exclusion, moving stochastically on a fluctuating landscape. An external field (for instance gravity) drives particles ($H$ and $L$) towards valleys of the landscape so as to minimize the overall energy, with the heavier $H$ particles preferentially displacing the lighter $L$'s upward. Moreover, $H$ and $L$ particles affect the evolution of the landscape in different ways.

As described in [4] and the companion paper [5], an interesting phase diagram is obtained as particle-landscape couplings are varied, with several distinct ordered phases arising. While there is a complete segregation of $H$ and $L$ particles in all phases, on a macroscopic scale, the spatial organization and movement of the landscape is different from phase to phase. This is reflected in the nomenclature of these phases: SPS (Strong phase separation), IPS (Infinitesimal current with phase separation) and FPS (Finite current with phase separation). In the SPS phase, the landscape is stationary, and forms a single macroscopic \/-shaped valley and a single /\ shaped hill, with $H$'s in the valley, and $L$'s on the hill. In the IPS phase, the current leads to the entire landscape falling with a velocity inversely proportional to the system size. While the $H$'s continue to occupy a deep \/ shaped valley, the landscape beneath the $L$-cluster shows a parabolic height profile. In the FPS phase, the fall velocity is finite. Although the $H$'s occupy a macroscopic valley, it now has a flatter, more rugged bottom, while the remaining part of the landscape with the $L$-cluster is disordered in this case. Average profiles of the landscape and particle occupancies are depicted in Fig. 1.

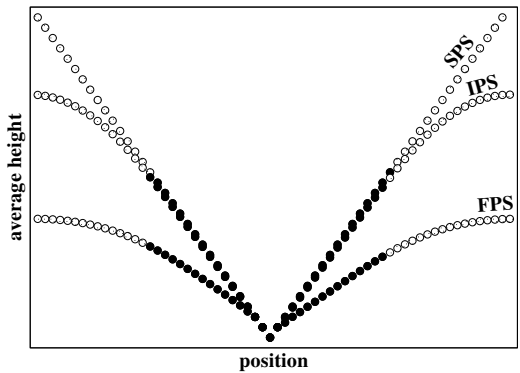

FIG. 1. Schematic description of various ordered phases. We show the average height profile of the landscape, $H$ ($L$) particles are shown by solid (empty) circles. In the SPS phase, due to complete phase separation between upslope and downslope bonds, average height grows linearly from the deepest point, with a slope 1. In the IPS phase, the height profile remains similar to SPS till the edge of the $H$ cluster and then it gradually flattens out. In the FPS phase, owing to partial phase separation between upslope and downslope bonds in the $H$-region, the height grows linearly with a slope smaller than unity, before flattening out in the $L$-region.

The detailed characterization of the static properties of these phases has been discussed in the companion (preceding) paper [5]. In this paper, we discuss the dynamic properties of the phases highlighting the basic point that the different types of ordering of the landscape result in very different dependence of time-scales on the system size $N$. Thus, in the SPS phase, motion of the interface between the coexisting phases involve an 'ergodic' time-scale, growing exponentially with $N$, so that there is no perceptible large-scale movement of rearrangement of the landscape on shorter timescales. By contrast, in the IPS and FPS phases, the coexistence of ordered and disordered landscape phases gives rise to novel steady state dynamics both near the interfaces and close to the bottom of the large valley, on time-scales which grow algebraically with $N$. To characterize this dynamics, in this paper, we propose a scaling ansatz and show that our simulation data for various dynamical correlation functions can indeed be described well by this ansatz. We also estimate different scaling exponents that summarize the scaling behavior of each correlation function considered here. In [4] it was noted that the properties of the landscape in the $L$-region has certain similarities with open systems. In this paper, we examine this issue further and show that there is a quantitative matching between the dynamics of the landscape in the $L$-region with that of an open-chain symmetric (asymmetric) exclusion process for the IPS (FPS) phase.

The rest of the paper is organized as follows. In the next section we present the model, briefly describe the different ordered phases and summarize the new results for each phase. In section III we present the results on the steady state dynamics of the landscape in the $H$ region, in IPS and FPS phases. In section IV we discuss dynamical properties of the landscape in the $L$ region. Section V contains results on the dynamics of the $H$ cluster. In section VI we present our results for the two-dimensional system. Our conclusions appear in section VII.

## II. THE $LH$ MODEL AND ITS PHASES

The $LH$ model describes two sets of hard-core particles sliding under gravity on a fluctuating landscape: (a) a set of heavier ($H$) particles which tend to minimize their energies by sliding down along the local height gradient towards local minima in the landscape, and further aid minimization of energy by pushing the landscape down under their own weight, and (b) a set of lighter ($L$) particles which are displaced upward by down-moving $H$ particles, along local height gradients and which affect the landscape around their positions by favoring certain transitions, thereby imparting a push to the landscape; as the coupling constant is varied, the push may be upward, nil, or downward. Each case results in a different phase. The model system describes the evolution of two coupled conserved fields, namely, $H$ or $L$ particle density, and the height gradient of the landscape, as microscopic moves involve exchange of $H$ and $L$ particles, or successive tilt variables which represent local slopes of the landscape, as described below.

We study the $LH$ model on a one-dimensional periodic lattice of size $N$. Each site holds either an $H$ or an $L$; there is only one particle per site, owing to hard-core exclusion between them. The lattice bonds on the other hand, are discrete surface elements that can assume two possible orientations with slopes $\tau_{i+1/2} = \pm 1$. The $H$ and $L$ particles preferentially interchange locations if the intermediate bond favours the $H$ to slide down. While the parts of the landscape rich in $H$-particles always tend to get pushed down, the parts occupied by $L$'s might get pushed up, pushed down at a rate slower or faster compared to the $H$'s, or fluctuate symmetrically with no effect of the $L$'s on them.

We explain our model schematically in Fig. 2. The solid black dots (hollow circles) denote the $H(L)$ particles while the symbols '/' and '\' correspond to up-slope ($\tau_{i+1/2} = 1$) and down-slope ($\tau_{i+1/2} = -1$) bonds, respectively. The dynamics conserves the total number of $H$ and $L$ particles on the lattice and also the total number of up and down-slopes. We consider an untilted surface with equal number of upslope and downslope bonds so that $\sum_i \tau_{i+1/2} = 0$.

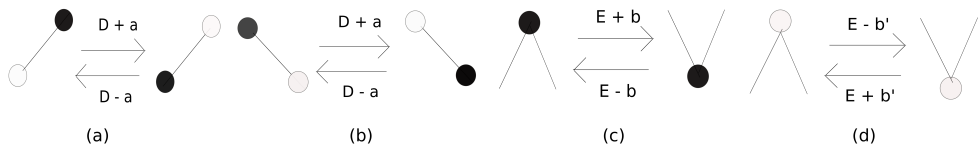

FIG. 2. The different transitions in the model and the rates for each of them. (a) and (b) show the rates of particle movements. An $H(L)$-particle slides down(up) with a rate $D + a$, while the reverse moves occur with a rate $D - a$, where $D > a > 0$. (b) and (c) show the rates at which the flipping of local hills and valleys occupied by respectively $H$ and $L$ particles takes place. A local hill (valley) occupied by an $H$ flips with a rate $E + b$ ($E - b$), where $E > b > 0$. A local hill valley), occupied with an $L$ on the other hand, can flip with rate $E - b'$ ($E + b'$) where the parameter $b'$ can be either zero, or positive or negative such that $E > |b'|$.

Our model shows three distinct well-ordered phases as the parameters $b$ and $b'$ are varied [4], corresponding to whether $L$ particles tend to impart an upward push, no push, or a downward push to the fluctuating surface. In [5] we have presented a detailed characterization of the static properties of these ordered phases. In this paper, we present results on their dynamical properties. Before doing so, we briefly summarize the properties of each ordered phase here.

**Strong phase separation** (SPS): $b, b' > 0$ corresponds to the SPS phase where the upslope and downslope bonds in the landscape completely phase separate from each other and form a deep $V$-shaped valley, as shown in Fig. 1.

A compact, macroscopic $H$-cluster is present at the bottom of the valley, while the $L$-cluster straddles the hill. The model with $b, b' > 0$ was studied earlier [6, 7] in the context of sedimenting colloidal crystals. In the SPS phase, separation of $H$ and $L$ particles as well as uptilts and downtilts is complete. Coarsening towards such a phase is a logarithmically slow process which can be explained as follows. In an intermediate state of relaxation, the system consists of a few deep valleys, each containing an $H$ cluster, separated by an equal number of hills containing the $L$-clusters. Since with $b, b' > 0$, the $H$ particles push the valleys downward, and the $L$ particles push the hills upward. The relaxation then involves an Arrhenius process across a large barrier, and the system gets stuck in metastable states [7]. As a result, the relaxation time grows exponentially with the system size in an SPS phase. Such diverging timescales allow for little dynamics in steady state. We do not study SPS dynamics further in this paper.

**Infinitesimal current with phase separation** (IPS): For $b > 0$, $b' = 0$ the $H$'s push the landscape down while the parts of the landscape occupied by "neutral" $L$'s undergo symmetric Edwards Wilkinson type fluctuations [8]. In the steady state, the $H$ and $L$ species undergo complete phase separation as in the SPS phase. But the average height profile of the landscape is different from SPS phase. A large valley is still formed here, but pure domains of up- and down-slope bonds extend from the bottom of the valley till the edges of the $H$-cluster, and not beyond that. The landscape beneath the $L$-cluster has no long-range order and the average height profile in this region assumes a parabolic shape, as shown in Fig. 1. The relaxation to this phase is algebraically fast as the symmetric fluctuations in the $L$ region make the merging of neighbouring valleys a diffusive process.

In the steady state, there is a current of upslope and downslope bonds that scales as $1/N$ for a lattice of size $N$. As a result of this current, the landscape, along with the particle clusters shows a downward motion with an average velocity $1/N$. In [4] this phase was named as "infinitesimal fall with phase separation" since for very large $N$, the velocity of the downward fall becomes infinitesimally small. The coexistence of an ordered valley along with the symmetrically fluctuating part of the landscape, gives rise to rich steady state dynamics with time-scales growing algebraically with $N$ for the landscape, and exponentially with $N$ for the particles.

In this paper, we present a numerical study of valley dynamics in the $H$ region by monitoring fluctuations both in the height and in the transverse direction. Results support a scaling description. Numerical evidence is also gathered for the valley bottom being confined by a quadratic potential, supporting the assertion that the dynamics is described by an Ornstein-Uhlenbeck process. In the $L$-region we provide detailed evidence for the description as an open SEP by monitoring the auto-correlation function as well as the mean-squared displacement of the tagged tilt. We also present results for two dimensions and show that while the valley dynamics in the $H$ region shows diffusive behavior, in the $L$-region the fluctuations in the two dimensional landscape grow logarithmically with time.

**Finite current with phase separation** (FPS): For $b > 0$, $-b < b' < 0$, the $L$'s push the landscape downward but at a lower rate than the $H$'s. In this phase, the upslope and downslope bonds of the landscape undergo partial phase separation such that a macroscopic valley is formed, which is much shallower than the valley obtained in IPS phase (see Fig. 1). This valley contains the $H$-cluster. The landscape beneath the $L$-cluster is disordered and the average height flattens out in this region, as shown in Fig. 1. The entire system carries a finite current of upslope and downslope bonds in the steady state resulting in a net downward motion with finite velocity. The fact that unlike IPS phase, the velocity of the downward fall remains finite in this case even for large systems, explains the name 'fast fall with phase separation' used in [4].

Here, we study and contrast the behavior of FPS valleys with their IPS counterparts. Although the valleys are more rugged in this case, a scaling description is still valid with important differences in exponents. Further, numerical evidences point to the fact that an Ornstein-Uhlenbeck type description is not possible in this case. In the $L$-region a study of sliding tag correlation functions provides evidence that the system is well-described by the open ASEP in the maximal current phase. In two dimensions, we find that the landscape dynamics in the $H$-region shows a sub-diffusive behavior, and in the $L$-region the landscape shows diffusive fluctuations. We also present numerical evidence for the movement of the centre of mass of the $H$ cluster on a time scale which grows exponentially with system size, for both IPS and FPS phases.

## III. LANDSCAPE DYNAMICS IN THE $H$ REGION

We are interested in characterizing the dynamics of the quintessentially nonequilibrium IPS and FPS phases in the steady state. To this end, we measure the mean-squared fluctuations

$$W(t, N) = \langle [Y(t) - Y(0)]^2 \rangle \tag{1}$$

as a function of time $t$, where $Y(t)$ is a generic stochastic variable. As discussed below, it may denote the position of the deepest point of the valley; or width of the $H - L$ domain wall; or height of the centre of $H$-cluster, etc. For each such quantity, our data show that $W(t, N)$ grows algebraically with time for small $t$, with a possible $N$-dependent co-efficient while for large $t$, it saturates to an $N$-dependent value. More precisely, $W(t, N) \sim t^\alpha / N^\nu$ for small $t$, and $\sim N^{\gamma}$ when $t$ is large. The values of the exponents $\alpha$, $\nu$ and $\gamma$ of course depend on the particular physical quantity $Y(t)$ represents, but a single scaling form suffices to describe all the results:

$$W(t, N) \sim N^{\gamma} F(t/N^z), \qquad (2)$$

The short and long time behavior of $W$ imply that the scaling function $F(x) \sim x^{\alpha}$ for small $x$ and saturates to a constant at large $x$ with exponents that are related through $z = (\gamma + \nu)/\alpha$. In an earlier work [4] we had measured the dynamical exponent which describes the coarsening during approach to the steady state, and had obtained $z = 2$ for both IPS and FPS phases. In the IPS phase, the same value of $z$ describes steady state dynamics as well. However, in the FPS phase, while the typical time-scale for relaxation towards the steady state $\sim N^2$, the typical time-scale for steady state dynamics $\sim N^{3/2}$.

Below we present our data for different dynamical correlations in the two phases.

## A. IPS phase

### 1. Valley dynamics

Let $C$ denote the centre of mass of the $H$ cluster, and let $V$ denote the deepest point of the valley which holds the $H$ cluster (see Fig. 3). The tilt current flowing in the $x$ direction through the system in steady state results in movement of $V$, whose position does not always coincide with $C$, since the latter is practically stationary and can move only over ergodic time-scale $\sim \exp(bN)$. To monitor the motion of $V$, we consider the case when $Y(t)$ in Eq. 1 denotes the position of $V$, and the mean-squared fluctuation of this position is denoted as $\sigma_0^2(t)$. Our earlier results in [4] show that $\sigma_0^2(t)$ grows diffusively with time with a diffusion constant $\sim 1/N$ for short times $t \ll N^2$, then saturates at a value $\sim N$ for large times $N^2 \ll t \ll exp(bN)$. This means that $V$ diffuses around this mean position $C$, but stays confined within a region of size $\sim \sqrt{N}$, and this region is explored by $V$ over an algebraic time-scale $\sim N^2$. In other words, we find $z = 2$, $\alpha = \nu = \gamma = 1$ in the scaling form with a linear growth of the scaling function close to the origin. The data presented in Fig. 4a gives evidence for this conclusion.

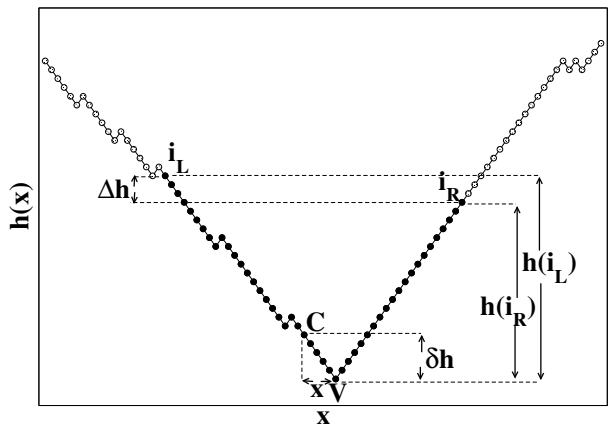

FIG. 3. The deepest point $V$ of the valley moves a distance $x$ from the centre of mass $C$ of the $H$ cluster. This causes a height asymmetry $\Delta h$.

The $1/N$ scaling of the short time diffusivity can be explained using a simple argument. At short times, the mechanism for valley dynamics involves the propagation of a downslope (upslope) bond through a pure domain of upslope (downslope) bonds. For example, if a downslope (upslope) bond reaches the upper edge of the $H$ particle cluster, then it zips ballistically through the domain of pure upslopes (downslopes) beneath the cluster and causes $V$ to shift by one unit towards the right (left). The time-scale associated with this process is proportional to $N$. Further, to estimate the time-scale for formation of a local hill at the edge of the $H$ cluster, we show in Appendix A that the average distance of the nearest downslope bond from the $H - L$ domain boundary scales as $\sqrt{N}$. This downslope bond reaches the $H$-cluster boundary diffusively at times $\sim N$. Thus the time-scale of the two relevant processes — formation of a local hill at the edge of the $H$-cluster, and transport of that hill towards the bottom of the valley — are both of order $N$. Since the processes occur with equal weight on the left and right arms of the valley, the motion of $V$ is diffusive, with diffusivity $\sim 1/N$.

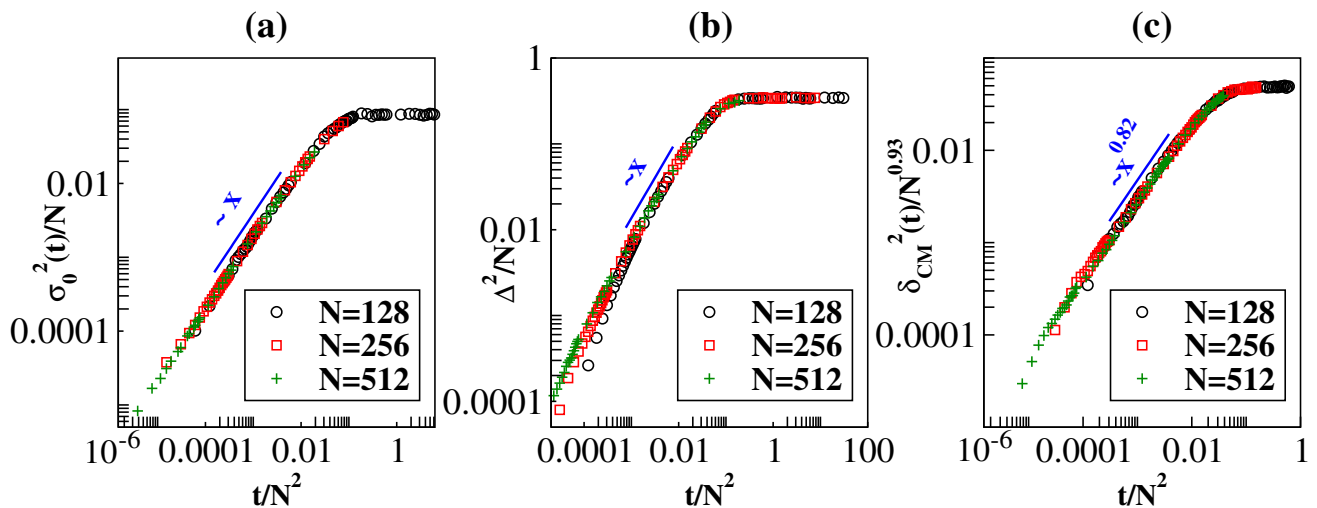

FIG. 4. Scaling collapse of dynamical correlation functions in IPS phase. The scaling argument is $t/N^z$ with dynamical exponent $z = 2$. For small argument, the scaling function grows as a power law with an exponent $\alpha$. (a): For $\sigma_0^2(t)$ we find $\gamma = 1$, $\alpha = 1$. (b): Scaling collapse of $\Delta^2(t)$ shows $\gamma = 1$, $\alpha = 1$. (c): For $\delta_{CM}^2(t)$ best collapse is obtained for $\gamma = 0.94$ and $\alpha = 0.82$. The maximum error bar in these exponent values are $\pm 0.02$. All data are averaged over at least $10^4$ initial histories.

### 2. Height fluctuations induced by valley dynamics

The deepest point $V$ of the valley which holds the $H$-cluster moves around $C$, the centre of mass of the cluster. If we associate a gravitational energy with the $H$-cluster, then it is easy to see that, for a clean V-shaped valley, the energy of the cluster is minimum when $V$ and $C$ coincide. Let $i_L$ and $i_R$ be the position of the left and right boundary of the $H$-cluster and $h(i_L)$ and $h(i_R)$ be the height of these two points measured with respect to $V$ (see Fig. 3). When $V$ is at $C$, we have $h(i_L) = h(i_R)$ and the energy is minimum. But when $V$ moves from this position, there is a height difference $\Delta h = h(i_L) - h(i_R)$, as shown in Fig. 3. The corresponding cost in the gravitational energy is proportional to $(\Delta h)^2$. Due to this energy cost, $V$ stays confined in a certain region around $C$. To demonstrate this, we monitor $\Delta^2(t) = \langle [\Delta h(t) - \Delta h(0)]^2 \rangle$, as a function of time. Here, $\Delta h(t)$ denotes the value of $\Delta h$ at time $t$.

We present our data in Fig. 4b. We find $\Delta^2(t)$ shows scaling behavior similar to $\sigma_0^2(t)$. This is expected, since the fluctuations in $\Delta h(t)$ are actually caused by movement of the valley around its mean position. An alternative way to probe the height fluctuations in the $H$-region is by measuring the fluctuations in the height difference $\delta h(t)$ between $C$ and $V$ (see Fig. 3). Its fluctuations are measured by $\delta_{CM}(t)^2 = \langle [\delta h(t) - \delta h(0)]^2 \rangle$. Although we expect similar scaling behavior for this quantity also, our data show slightly different values of the exponent. We find $\alpha = 0.82 \pm 0.01$, $\nu = 0.7 \pm 0.01$ and $\gamma = 0.94 \pm 0.02$, which yields $z = 2$. We verify this scaling form in Fig. 4c.

### 3. Valley dynamics as an Ornstein-Uhlenbeck process

In the previous subsection we saw that any displacement of $V$ away from $C$ costs energy, which in turn confines $V$ in a certain region around $C$. In this section, we show that the restoring force experienced by $V$ is simple harmonic in nature and in a continuum description, the diffusive motion of $V$ can therefore be described by an Ornstein-Uhlenbeck process [9], for which the Fokker-Planck equation has the form

$$\partial_t P(x,t) = \frac{1}{\tau} \partial_x [x P(x,t)] + D \partial_x^2 P(x,t) \qquad (3)$$

where $P(x,t)$ is the probability to find $V$ at a distance $x$ away from $C$. The first term on the right hand side represents the drift which can be positive or negative depending on the sign of $x$. The parameter $\tau$ sets the time-scale which is related to the steepness of the simple harmonic potential. We directly measure the drift term by measuring the bias experienced by $V$ at a position $x$. In our lattice model, this bias is defined as the difference between the rightward and leftward hopping rate from that position. For a given value of $x$, we measure the average waiting times $T^{\pm}(x)$ for $V$ to move to position $(x \pm 1)$. The bias is given by $v(x) = [1/T^+(x) - 1/T^-(x)]$. In Fig. 5a we plot the bias as a function of $x$ and find that its magnitude increases linearly. This finding establishes the simple harmonic nature of the force field.

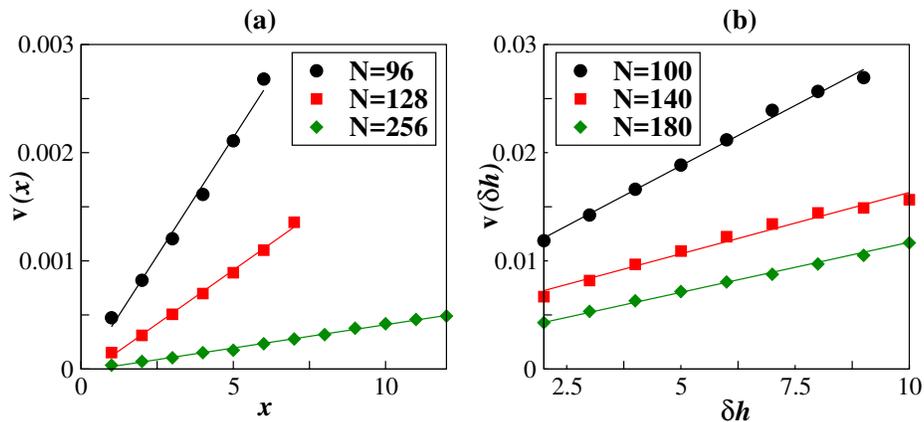

FIG. 5. Motion of $V$ in IPS phase as an Ornstein-Uhlenbech process. (a): Velocity of $V$, when at a distance $x$ away from $C$, increases linearly with $x$. (b): $C$ experiences a restoring bias, when at a height $\delta h$ above $V$ and the bias increases linearly with $\delta h$. The data shown for 3 different system sizes have been averaged over at least $10^4$ initial histories.

An alternative description of the above dynamics can also be given in terms of the random variable $\delta h$ denoting the height difference between $V$ and $C$. Note that by definition, $\delta h$ can never cross the origin and become negative. In a single time-step, the value of $\delta h$ may change by one unit, or even two units (when the centre of mass is on a local hill which flips). We measure the average waiting times $T_n^\pm(\delta h)$ till $\delta h$ changes by $\pm n$ units, with $n = 1, 2$ and calculate $V(\delta h) = \sum_{n=1,2}[n/T_n^+(\delta h) - n/T_n^-(\delta h)]$. We find that $V(\delta h)$ increases linearly with $\delta h$ (see Fig. 5b). This data show that the motion of $\delta h$ can be described as an Ornstein-Uhlenbeck process with reflecting boundary condition at the origin.

### B. FPS phase

#### 1. Valley dynamics

For the FPS phase, the valley dynamics shows a different behavior from the IPS phase. Although the deepest point $V$ of the valley still makes excursions around the centre of mass $C$ of the $H$-cluster, this motion is not diffusive as in the IPS phase. Our data show that $\sigma_0^2(t)$ grows sub-diffusively $\sim t^{1/2}$. The sub-diffusive regime is observed after a large non-scaling regime but before the saturation sets in, and is observable provided $N$ is large enough. Our data show $\alpha = 0.5 \pm 0.01$, $\nu = -0.17 \pm 0.02$, $\gamma = 0.92 \pm 0.03$, which yields $z = 1.5$ and a scaling function which grows with an exponent $1/2$ close to the origin. The corresponding scaling collapse is shown in Fig. 6a.

Note that in the FPS phase, the valley is not as sharp as the one seen in IPS phase. Instead of a single deepest point, the bottom of the valley is often rugged and may contain more than one site with the minimum height. In our simulation, we have chosen one such site and every time a random update causes the height of a site to fall below the current minimum, we count it as a displacement of the deepest point. To verify that the value of the exponent is not tied to the particular method of determining it, we also estimate the exponent by monitoring the time dependence of the mean-squared displacement of a tagged upslope bond that is situated at a distance $r$ away from the centre of mass in the $H$ region. As shown in appendix C, this quantity also grows with the exponent $1/2$, as did $\sigma_0^2(t)$.

An important qualitative difference in the scaling form in IPS and FPS phase is the sign of the exponent $\nu$. While $\nu$ is positive for IPS phase, in the FPS phase we find $\nu < 0$. This implies that as $N$ becomes larger, the growth of $\sigma_0^2(t)$ with time becomes slower for IPS phase, and faster for FPS phase. This difference can be related to the difference in the current fluctuation properties. Note that for a general current-carrying system, in which the state breaks translational invariance, the current fluctuations in steady state can be very different in parts of the system, though the mean current is uniform. Indeed we find the current fluctuation at the bottom of the valley is very different from that measured in the middle of the $L$-phase. The exponent $\nu$ is related to the fluctuation properties at the bottom of the valley. We find that (data not shown here) for an FPS system, which carries a finite current, the current fluctuation at the valley bottom grows with $N$. By contrast in the IPS phase, where current $\sim 1/N$, its fluctuations get smaller as $N$ increases. This ties in with the different signs of $\nu$ in the two phases.

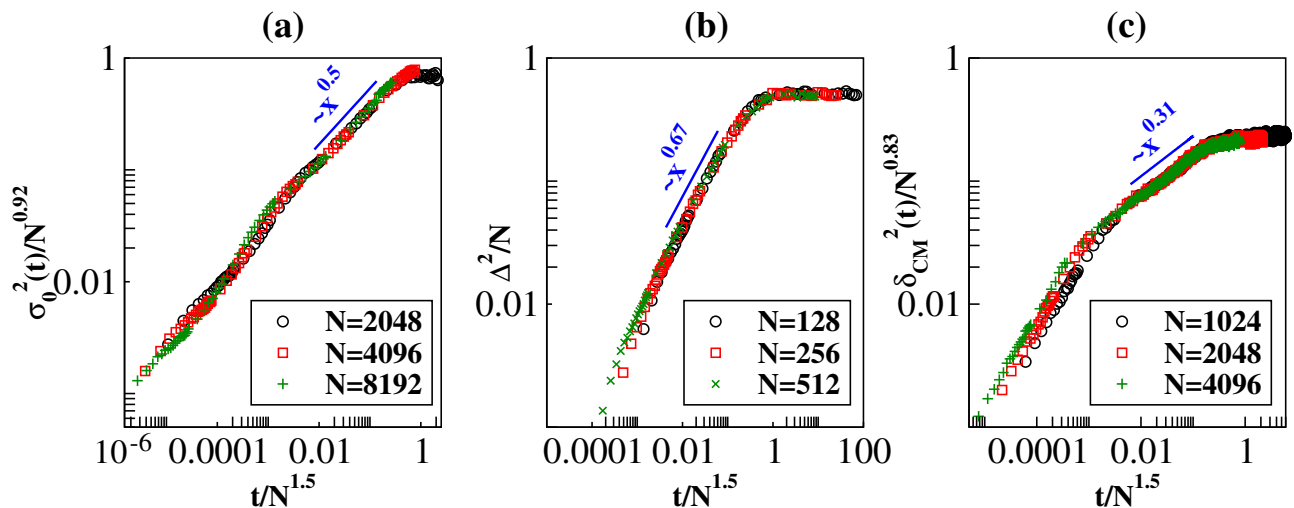

FIG. 6. Scaling collapse of dynamical correlation functions in FPS phase. In this case the scaling argument is $t/N^{3/2}$. For small argument the scaling function grows with a power $\alpha$ and for large argument saturates to a constant value. (a): Data for $\sigma_0^2(t)$ give $\gamma = 0.92$ and $\alpha = 0.5$. (b): For $\Delta^2(t)$ we find $\gamma = 1$ and $\alpha = 0.67$. (c): For $\delta_{CM}^2(t)$ best collapse is obtained for $\gamma = 0.83$ and $\alpha = 0.31$. The maximum error bar is $\pm 0.02$. All these data have been averaged over at least $10^4$ steady state configurations.

### 2. Height fluctuations induced by valley dynamics

In the FPS phase, $\Delta^2(t)$ grows sub-diffusively with time with an exponent $\simeq 0.67$ (Fig. 6b). This value is close to the value expected for KPZ-type fluctuations. Actually, fluctuations originating in the $L$-region of the system can cause fluctuations in $\Delta(t)$, and our discussion in section IV B shows that in $L$-region the fluctuations indeed are of KPZ type. For another related quantity $\delta_{CM}^2(t)$ that measures the fluctuations in $\delta h$ (defined in Fig. 3), we find the growth exponent $\alpha = 0.31 \pm 0.02$ (Fig. 6c). Other exponents characterizing the behavior of $\delta_{CM}^2(t)$ are $\nu = -0.36 \pm 0.02$, $\gamma = 0.83 \pm 0.02$.

Note that $\sigma_0^2(t)$, $\Delta^2(t)$ and $\delta_{CM}^2(t)$ grow with different exponents at short times. Although dynamics of $V$ affects $\delta h$ or $\Delta h$ (see Fig. 3), due to the rugged structure at the bottom of the valley in FPS phase, there is no one to one correspondence between the dynamics of these quantities. In Fig. 7 we have shown a specific example where $V$ moves but $\Delta h$ does not change. As a result, the corresponding dynamical correlation functions show different scaling exponents. However, apart from the KPZ-type growth for $\Delta^2(t)$, mentioned in the previous paragraph, we do not have clear understanding of the quantitative values of the scaling exponents observed for $\sigma_0^2(t)$ and $\delta_{CM}^2(t)$ at this stage.

In the FPS phase, the valley motion cannot be described in terms of an Ornstein-Uhlenbeck process. This is not surprising, in view of the sub-diffusive nature of the valley dynamics (an underlying Ornstein-Uhlenbeck process would have yielded a diffusive motion of the deepest point). Indeed Fig. 8 shows that the bias experienced by $V$ when it is at a height $\delta h$ below $C$, grows sub-linearly with $\delta h$.

## IV. LANDSCAPE DYNAMICS IN THE $L$-REGION

In the previous subsection, we discussed the dynamics of the ordered part of the landscape, in the region occupied by $H$ particles. The landscape beneath the $L$-cluster is disordered in both the IPS and FPS phases. In this part the dynamics of the landscape actually corresponds to that of an open-chain exclusion process, on mapping the upslope bonds to particles and the downslope bonds to holes.

### A. IPS phase

For an IPS phase, in the language of particles and holes, in the $L$-region the dynamical rules (see Fig. 2c) are those of a symmetric exclusion process (SEP). The ordered domains of upslope and downslope bonds in the $H$-region (also see Fig. 1) then act as reservoirs of particles and holes that are present at the two ends of the macroscopic SEP segment, and drive a current through it. Earlier studies on an open-chain SEP [10] show that the density profile of particles shows a linear gradient $\sim 1/N$ which supports a diffusive current $\sim 1/N$ in the system. We had verified both these predictions in earlier work [4].

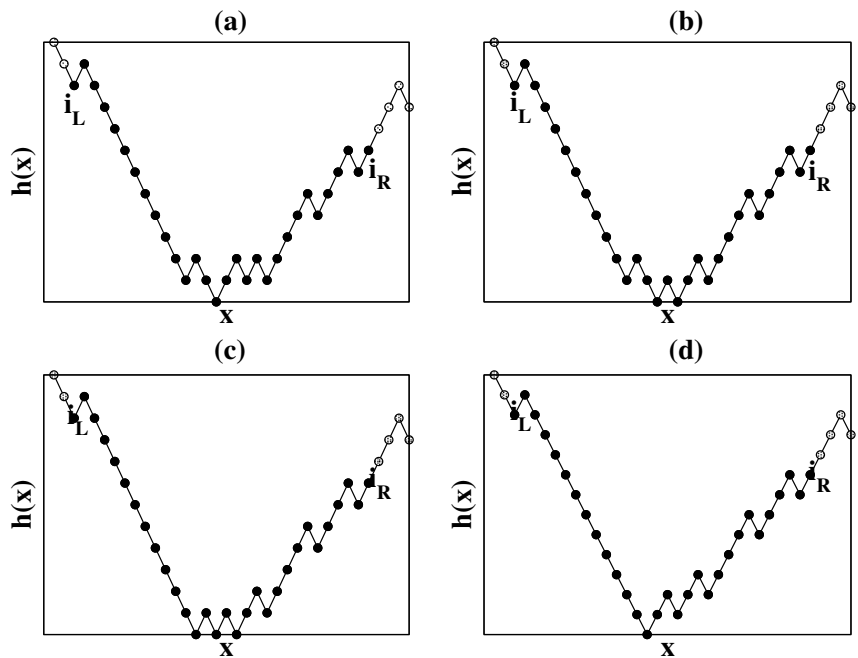

FIG. 7. We explicitly illustrate that it is possible for $V$ in an FPS phase to move without causing any change in the height difference between the edges of the $H$-cluster. Figs (a), (b), (c), (d) show time-evolution of a local configuration in the FPS phase (also see the average structure of the phase in Fig. 1). We find that the finite fraction of impurities present within the majority up(down)-slope domains causes the deepest point to shift without causing any change in the heights of the right ($i_R$) and left($i_L$) edges of the $H$ cluster.

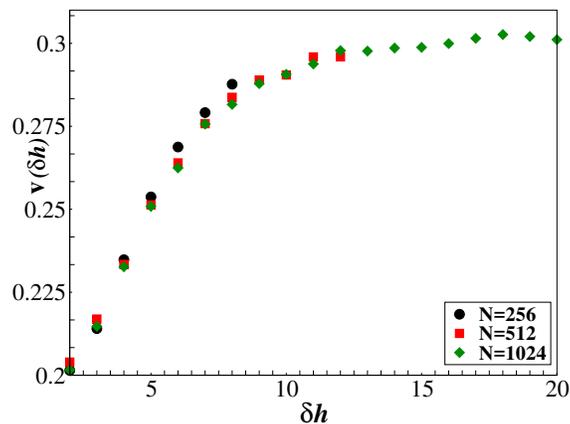

FIG. 8. Velocity of $V$ when at a height $\delta h$ below $C$ has a non-linear dependence on $\delta h$ in the FPS phase. This motion cannot be described as an Ornstein-Uhlenbeck process, unlike IPS phase. The data shown for 3 different system sizes have been averaged over at least $10^4$ histories.

In this paper, we examine the correspondence with open system at a more quantitative level. We compare the dynamical correlation functions in the $L$-region and compare it with the exact calculation in SEP. The auto-correlation function of an upslope bond at a distance $r$ away from the $H-L$ domain boundary is shown in Fig. 9a. The continuous line in this plot shows the result expected for open-chain SEP, where $A(r,t) = \frac{\rho(r)(1-\rho(r))}{\sqrt{2\pi}} t^{-1/2}$, with $\rho(r)$ being the local density at a position $r$ [11]. We find a good agreement between this prediction and our simulation results. In Fig. 9b, we plot the mean-squared displacement of a tagged upslope bond at an initial distance $r$ from the $H-L$ domain boundary. This should be compared with the mean-squared displacement of a tagged particle in an open-chain SEP and is expected to behave as $\sigma^2(r,t) = \sqrt{\frac{2}{\pi}} \frac{1-\rho(r)}{\rho(r)} t^{1/2}$ [12], up to a time $t \sim N^2$. Beyond this time, the tagged particle displacement becomes larger than the length scale over which $\rho(r)$ varies, and $\sigma^2(r,t)$ then shows a faster growth. Our numerical data for tagged upslope displacement, shown by discrete points in Fig. 9b, match with tagged particle displacement in SEP (continuous line in Fig. 9b) for shorter times and shows deviation for large times, as argued above.

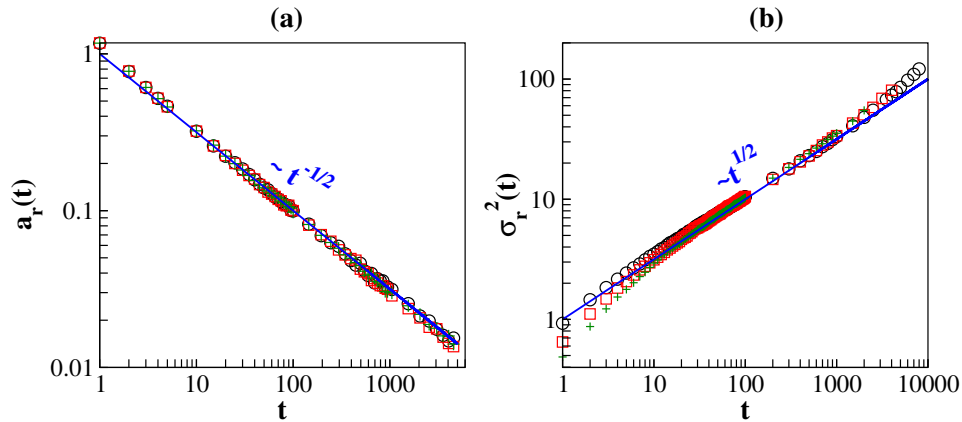

FIG. 9. The correspondence between the $L$ region in the IPS phase and an open-chain SEP. a: Scaled auto-correlation function of an upslope bond at a distance $r + L/4$ from $C$, where the average density of upslopes is denoted by $S^+(r)$. We plot $a_r(t) = \frac{\sqrt{2\pi}}{S^+(r)(1-S^+(r))} A(r,t)$ for 3 different $r$ and show that it decays as $t^{-1/2}$. The system size used is 2048. The data shown here have been averaged over at least $10^6$ initial histories. b: Tagged mean-squared displacement of an upslope bond at a distance $r + L/4$ from $C$. We plot $\sigma_r^2(t) = \sqrt{\frac{\pi}{2}} \frac{S^+(r)}{1-S^+(r)} \sigma^2(r,t)$ for 3 different values of $r$ and show that it grows as $t^{1/2}$. We have used $N = 512$ here. These data have been averaged over at least $10^4$ initial histories.

## B. FPS phase

In the FPS phase, a similar mapping of upslope (downslope) bonds to particles (holes) brings out a correspondence with an open-chain asymmetric exclusion process (ASEP) in the maximal current phase [13]. In [4], we have explicitly shown that the density of upslope bonds in the bulk of the $L$-region is $1/2$ and near the $H - L$ domain boundaries it shows an algebraic variation, as expected in a maximal current phase. Here, we extend the comparison to dynamical correlation functions. We measure the mean-squared displacement of a tagged upslope bond in the $L$-region. Note that due to the presence of a kinematic wave [14], the density fluctuations in an ASEP with density $\rho$ propagate with a velocity $(1 - 2\rho)$, which differs from the average particle velocity $(1 - \rho)$. To probe how a particular density fluctuation decays with time, one needs to measure the tagged mean-squared displacement with a 'sliding tag' that keeps changing with time (see Appendix B for a detailed explanation). In [15, 16] this method was employed for an ASEP with periodic boundary conditions. In this paper, we generalize this method for an open system and find that the tagged mean-squared displacement of an upslope bond grows with time with an exponent close to $2/3$, as expected for an ASEP (see Fig. 10).

## V. MOTION OF THE $H$-CLUSTER

In steady state, along with the rapid dynamics of the landscape, the particles show a very slow dynamics over a time-scale, which grows exponentially with system size. We measure the mean squared displacement ($\sigma_{CM}^2$) of the centre of mass of the $H$-cluster and find that it grows diffusively with a diffusivity $\sim exp(-\alpha N)$. The data for the IPS phase is shown in Fig. 11. A similar behaviour is found for the FPS phase too.

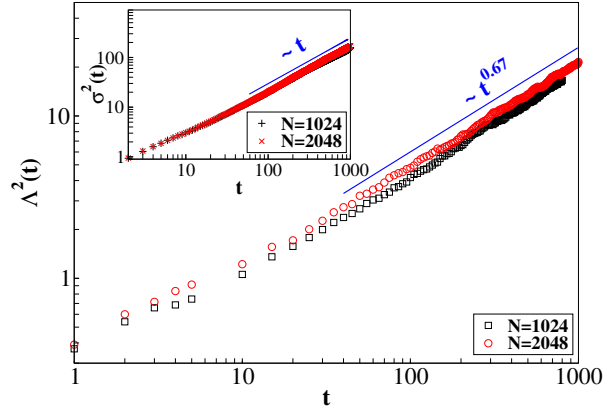

FIG. 10. Main plot shows the sliding tag correlation function $\Lambda^2(t)$ for an upslope bond in the $L$ region in FPS phase. It grows as $t^{0.67}$. Inset shows the variance $\sigma^2(t)$ of a tagged upslope bond displacement in the same phase. When the relative shift between the tagged bond and the density patches is not taken into account, measurement of the tagged mean squared displacement does not capture the correct dissipation and shows a diffusive behaviour. We have used $N = 1024, 2048$ here. We have taken the density of $H$ particles to be $1/2$ and hence the length of the open chain ASEP in maximal current phase is of length $N/2$. These data have been averaged over at least $10^3$ initial histories.

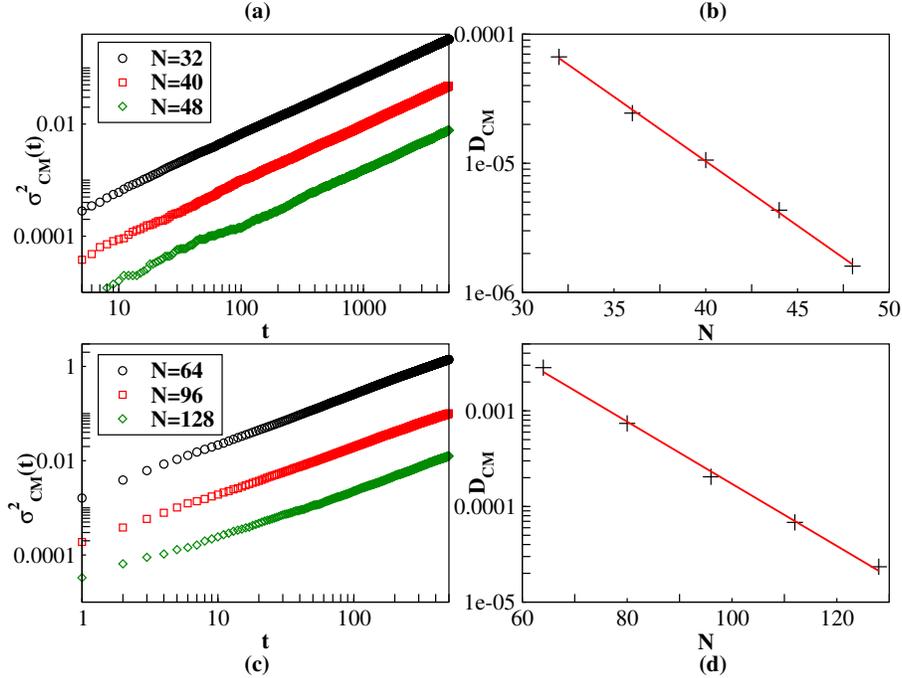

FIG. 11. Tagged mean squared displacement ($\sigma_{CM}^2$) of the centre of mass of the $H$-cluster for IPS and FPS phases. $\sigma_{CM}^2$ grows diffusively in time with a diffusivity $D_{CM} \sim exp(-\alpha N)$. Plots (a) and (b) show the behaviour in the IPS phase, while (c) and (d) correspond to the FPS phase. The values of $\alpha$ in IPS and FPS phases are $\sim 0.23$ and $\sim 0.1$ respectively. These data have been averaged over at least $10^5$ initial histories.

## VI. STEADY STATE DYNAMICS IN TWO DIMENSIONS

The model defined in Fig. 2 can be generalized in two dimensions. In this case, $h(i, j)$ denotes the height at a site $(i, j)$ on the two-dimensional square lattice. The height difference between the nearest neighbor sites can be $\pm 1$. A site $(i, j)$ is a local hill if all its four neighbors at $(i \pm 1, j)$ and $(i, j \pm 1)$ are at a height $h(i, j) - 1$. Similarly, the site is a local valley when all the neighboring sites have a height $h(i, j) + 1$. The transition rates between hills and valleys occupied by $H$ or $L$ particles remain same as in Fig. 2. The particles hop from one lattice site $(i, j)$ to any of its neighboring site and if the destination site has a lower height, then $H$ particles preferentially displace the $L$'s, as in

Fig. 2. We present our simulation results on the dynamical correlations in the two-dimensional model in this section.

Algebraically fast time scales in relaxation and steady state dynamics of the landscape are observed in the IPS and FPS phases in two dimensions as well. We monitor the mean squared displacement of the deepest point in the valley along the $x$ and $y$ directions in the IPS and FPS phases (Fig. 12a, b). We find that in the IPS phase, at short times $t \ll N$, the valley moves diffusively with the diffusion constant $\sim 1/N$. The mean squared displacement then saturates at a finite value of order 1 until at large times it again starts growing diffusively, but with a diffusion constant $\sim e^{-\alpha N}$. In the FPS phase, we monitor the same quantity and find that it grows sub-diffusively and saturates at a finite value. At large times $\sim e^{\alpha N}$, the valley performs diffusive motion with an exponentially small diffusion constant.

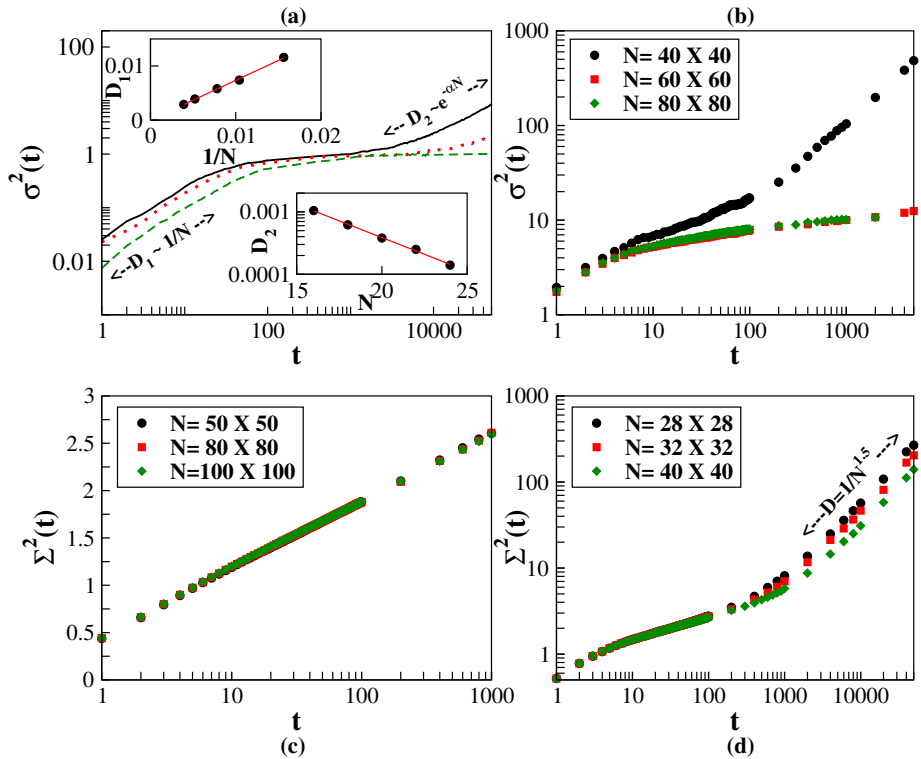

FIG. 12. Landscape dynamics in two dimensions. **(a)** Mean squared displacement $\sigma^2(t)$ of the deepest point of the valley along the $x$-direction as a function of time for three different system sizes, $24 \times 24$(solid line), $32 \times 32$(dotted line), $40 \times 40$(dashed line) in the IPS phase. Top and bottom insets show the short and large time diffusion constants $D_1 \sim 1/N$ and $D_2 \sim e^{-\alpha N}$ where $\alpha = 0.25$. **(b)** $\sigma^2(t)$ in the FPS phase for 4 different system sizes. While the initial growth and saturation does not show any dependence on the system size, the large time diffusivity falls off exponentially as the system size (data not shown here). **(c)** The height fluctuation $\Sigma^2(t)$ as a function of time in the part of the surface holding the $L$ cluster in IPS phase. The fluctuations grow logarithmically in time as is characteristic of an Edwards Wilkinson surface in two dimensions. **(d)** $\Sigma^2(t)$ in the FPS phase shows a sub-diffusive regime that remains valid upto $t \sim 100$. This is followed by a diffusive growth at large times with the diffusion constant falling off as a power law $\sim N^{-1.5}$.

We also measure height fluctuations $(\Sigma^2(t))$ in the part of the surface occupied by the $L$ particles and find that it shows a logarithmic growth for the IPS phase and a diffusive growth in the FPS phase with a diffusion constant $\sim N^{-1.5}$ (see Fig. 12c, d). Note that, while in the IPS phase, the surface beneath the $L$-cluster shows the logarithmic behaviour characteristic of an Edwards Wilkinson surface [17], in the FPS phase, there is a clear deviation from the behaviour of usual Kardar-Parisi-Zhang surfaces where one expects $\Sigma^2(t) \sim t^{2\beta}$ with $\beta \simeq 0.23$ [18].

## VII. CONCLUSION

Our study of the dynamical properties of a strongly coupled system of heavy and light particles on a fluctuating landscape has focused on the phase separated states that result when linear hydrodynamics predicts that the homogeneous system is unstable. As detailed in [5], there are several different phases which arise as coupling constants are changed. The dynamics of two such ordered phases (SPS and FDPO) have already been studied in [6, 7] and [19], respectively. In this paper, we have chosen to focus on the IPS and FPS phases. Their dynamic properties reflect some similarities and some differences.

Perhaps the single most striking feature in both phases is the existence of two very different time scales for the particles and the landscape in steady state, despite the strong coupling between them. Movement of the centre of mass of the macroscopic $H$ cluster occurs on a time scale which grows exponentially with system size, $\sim \exp(\alpha N)$, whereas landscape fluctuations involve power law growths up to a time $\sim N^\gamma$. This dichotomy can be traced to the fact that the disordered landscape beneath the $L$ cluster generates slope fluctuations, which reach the valley bottom and cause it to undulate, without a concomitant shift of the particles themselves. Interestingly, there is no such dichotomy in the coarsening properties of the systems. The approach to the steady state is described by power laws in time for both the IPS and FPS phases.

An interesting aspect of the coexisting phases in both cases is that their properties relate to well-known paradigms in the field, on employing a particle-hole description of the landscape slopes in the $L$ region. Thus, in the IPS we have effectively an open simple exclusion process (SEP), with boundary conditions which set up a gradient and cause a current to flow. In the FPS, the corresponding particle-hole model is the asymmetric simple exclusion process (ASEP), in the maximal current phase. The currents carried in both cases translate into a bodily downward movement of the interface, resulting in a velocity which is $\mathcal{O}(1/N)$ for the IPS and $\mathcal{O}(1)$ for the FPS.

Although we have attempted to be exhaustive and to rationalize the observed results to the extent possible, some mysteries remain. First, why is the dynamic exponent $z = 2$ in the coarsening regime of the FPS, whereas in steady state, it has the value $3/2$, as expected on the basis of the ASEP analogy? Secondly, why is $\nu$ negative in the FPS phase? We have pointed out a possible correlation between this and enhanced current fluctuations at the valley bottom, but it would be desirable to make this connection stronger.

## VIII. ACKNOWLEDGEMENTS

The computational facility used in this work was provided through the Thematic Unit of Excellence on Computational Materials Science, funded by Nanomission, Department of Science and Technology, India. M.B. acknowledges the award of the J. C. Bose National Fellowship by the Science and Education Research Board, India.

## Appendix A: Average distance of the nearest downslope bond from the $H - L$ domain boundary

In the part of the surface occupied by the $L$ particles, the density profile of the surface bonds shows a linear gradient [4]. In this appendix, within the mean-field approximation, we calculate the average separation between the first downslope bond and the $H - L$ domain boundary (which occurs at a distance of $N/4$ from the centre of mass of the $H$ cluster).

Let $P(r)$ be the probability that the first downslope bond is located at a distance $r$ from the $H-L$ domain boundary, i.e. there is an upslope bond for all $j < r$, and a downslope bond at $j = r$. Within the mean-field approximation, this probability is given by

$$P(r) = \prod_{j=1}^{r-1} \left(1 - \frac{2j}{N}\right) \frac{2r}{N} \tag{A-1}$$

The average value of $r$ is then given by

$$\langle r \rangle = \frac{2}{N} + \sum_{r=2}^{N/2} \sum_{j=1}^{r-1} \left(1 - \frac{2j}{N}\right) \frac{2r^2}{N} \tag{A-2}$$

To proceed further, define

$$z = \prod_{j=1}^{r-1} \left(1 - \frac{2j}{N}\right) \tag{A-3}$$

$$\log z = \frac{N}{2} \int_{2/N}^{2(r-1)/N} \log(1-y) dy \approx \frac{2r}{N} - \frac{r^2}{N} \tag{A-4}$$

Performing the integration and using this expression in Eq. A-2 one has, for large $N$,

$$< r > = \frac{2}{N} + (\sqrt{N} + 1/N)e^{1/N} \sim \sqrt{N} \tag{A-5}$$

### Appendix B: Details of measurement of sliding tag correlation $\Lambda^2(t)$

In the stationary state of an ASEP, the drift velocity ($v_p$) of the individual particles is given by $J/\rho$, where $J$, $\rho$ are particle current and density respectively. Besides this, the velocity ($v_k$) with which the coarse-grained density fluctuations are transported throughout the system is given by $\frac{\partial J}{\partial \rho}$ and is called the kinematic wave velocity [14]. Hence, in the rest frame of the density fluctuations, the particles move with velocity $v_p - v_k$. In order to capture the correct dynamical exponent with which a tagged particle in the maximal current phase dissipates away from its initial environment, we take recourse to the method of sliding tags [15, 16]. For an ASEP with open boundaries, due to finite rates of injection (ejection) at the boundary sites, the tags of the particles keep changing with time. To take this into account, one may consider a segment of length $l$ sufficiently away from the boundaries within the bulk and within this segment, tag all the particles at $t = 0$. Due to hard-core repulsion, none of these these can cross each other. We measure the following quantity:

$$\Lambda^2(t) = \langle [y(m', t) - y(m, 0)]^2 \rangle \qquad \text{(B-1)}$$

where, $m$, $m'$ are particle tags related by $m' = m - \rho u t$, $\{y(m, t)\}$ give the locations of the particles at time $t$ and $u = J/\rho$. Here, $J$ is the integrated current and $\rho = 1/2$ as the segment is in the maximal current phase. The angular brackets denote averaging over both initial histories and stochastic evolution. By carrying out this measurement until the tagged particles approach the boundaries of the segment, one finds $\Lambda^2(t) \sim t^{2/3}$.

### Appendix C: Mean squared displacement of a tagged upslope bond within $H$-cluster in FPS phase

In the $H$ region, we tag an upslope bond situated a distance $r$ away from the centre of mass and monitor its mean-squared displacement with time. Our data in Fig. C-1 shows that this quantity also grows with the same exponent $1/2$ as found for the mean squared displacement of the deepest point of the landscape.

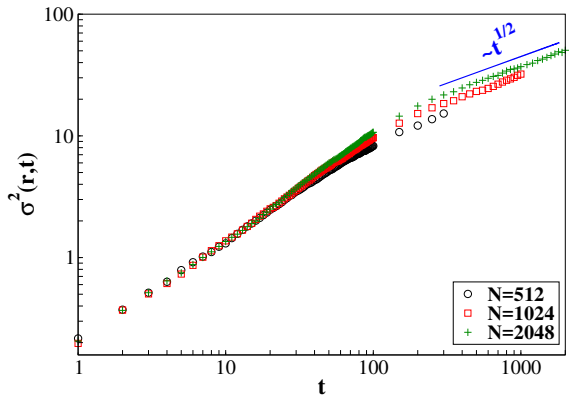

FIG. C-1. In FPS phase, the tagged mean squared displacement $\sigma^2(r, t)$ of an upslope bond inside the $H$-cluster initially at a distance $r$ from the deepest point shows a sub-diffusive behaviour and grows as $t^{1/2}$. Here, we tag an upslope bond at a distance $N/16$ from the deepest point and show the plots for 3 different $N$ values. We observe an initial non-scaling behavior for $t \lesssim 100$ where the growth is steeper. These data have been averaged over at least $10^5$ initial configurations.